\documentclass[conference]{IEEEtran}
\IEEEoverridecommandlockouts
\usepackage{cite}
\usepackage{amsmath,amssymb,amsfonts}

\usepackage{algorithmic}
\usepackage{graphicx}
\usepackage{textcomp}
\usepackage{xcolor}

\usepackage[ruled,vlined,linesnumbered]{algorithm2e}
\usepackage{subcaption}
\usepackage[hyphens]{url}
\usepackage[breaklinks=true]{hyperref}
\hypersetup{hidelinks=true, }
\usepackage[nameinlink]{cleveref}
\usepackage{xcolor, colortbl}
\usepackage{fontawesome5}
\usepackage{stmaryrd}
\usepackage{wrapfig}

\usepackage{cancel}
\newcommand{\sysname}{COoL-TEE\xspace}
\newcommand{\figpathmodel}{imports/img/SystemMulti-CloudModel.pdf}
\newcommand{\figpatharch}{imports/img/SystemOverview.pdf}
\newcommand{\figpathattack}{imports/img/AttackModel.pdf}
\newcommand{\figpathmotivation}{"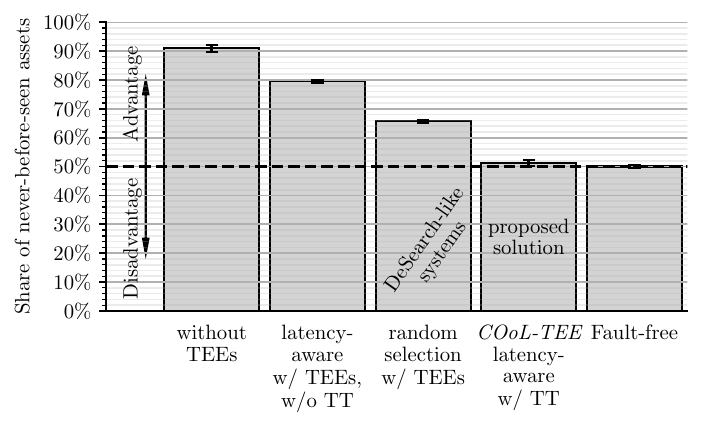"}
\newcommand{\figpathcasemulti}{"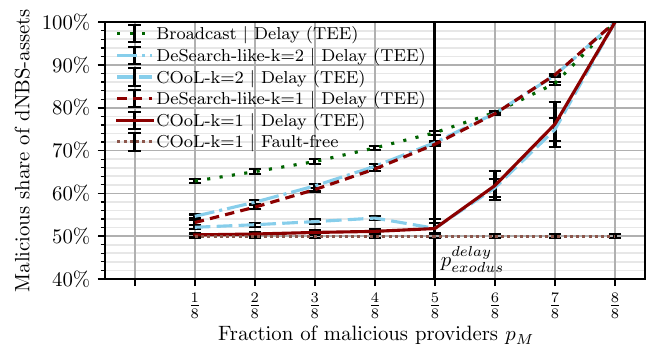"}
\newcommand{\figpathcaseba}{"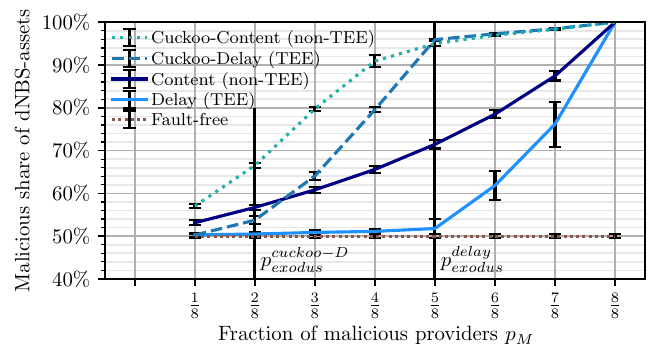"}
\newcommand{\figpathcasebb}{"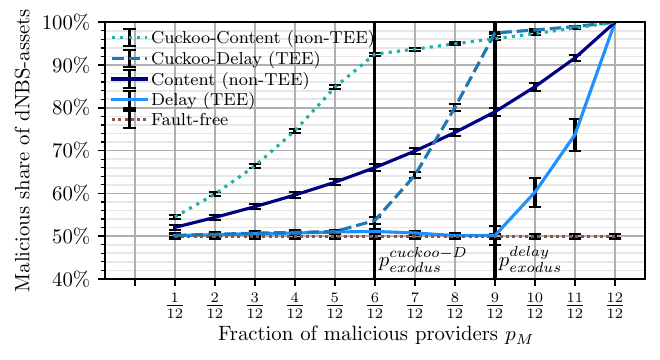"}
\newcommand{\figpathcasebTT}{"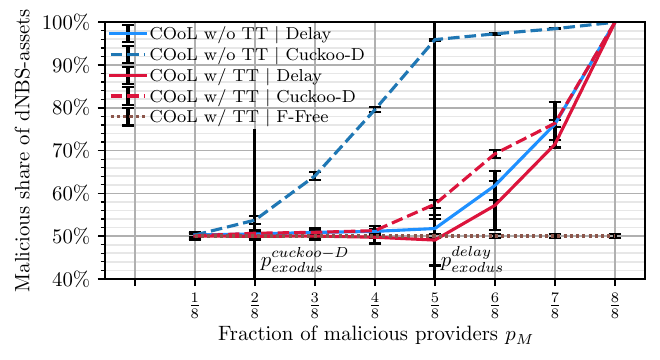"}
\newcommand{\figpathcasespot}{"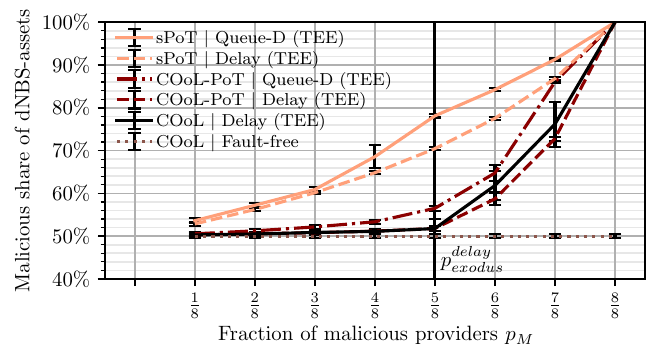"}
\newcommand{\figpathmdc}{"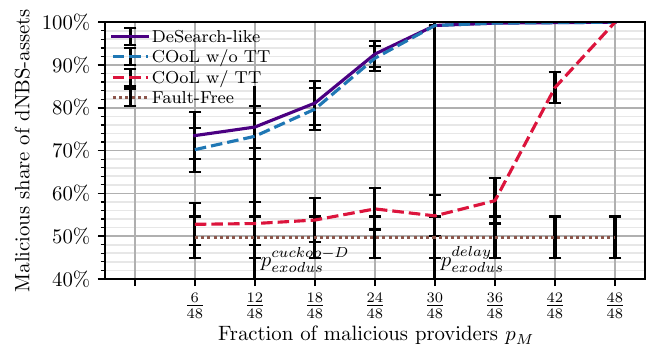"}
\newcommand{\figpathlatthrough}{"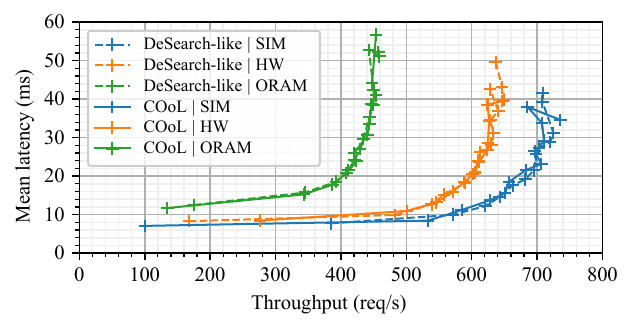"}
\newcommand{\figpathlatthroughTT}{"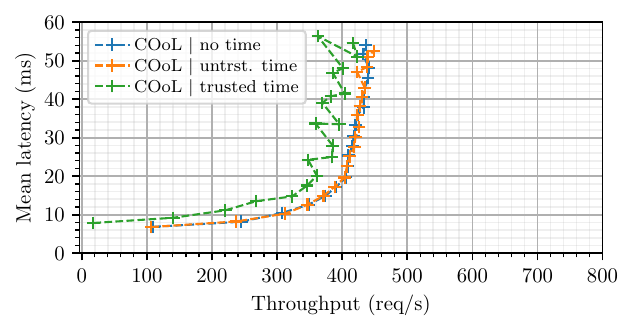"}

\newboolean{showcomments}
\setboolean{showcomments}{true}
\ifthenelse{\boolean{showcomments}}
{ \newcommand{\mynote}[3]{
        \fbox{\bfseries\sffamily\scriptsize#1}
        {\small\textsf{\emph{\color{#3}{#2}}}}}}
{ \newcommand{\mynote}[3]{}}
\newcommand{\awaitingreview}[1]{\faIcon{balance-scale}~}

\newcommand{\wontdo}[1]{\faIcon{times}~}

\def\BibTeX{{\rm B\kern-.05em{\sc i\kern-.025em b}\kern-.08em
    T\kern-.1667em\lower.7ex\hbox{E}\kern-.125emX}}
\begin{document}

\title{\sysname: Client-TEE Collaboration \\ for Resilient Distributed Search 
\thanks{This work was supported by a French government grant managed by the Agence Nationale de la Recherche under the France 2030 program, reference ``ANR-22-PEFT-0002'' as well as the ANR Labcom program, reference ``ANR-21-LCV1-0012''.
We thank the Complex Systems research group at the IIUN (University of Neuchâtel, Switzerland) for granting us access to their experimental platform.
}}

\newboolean{anon}
\setboolean{anon}{false}
\newboolean{oneline}
\setboolean{oneline}{true}
\ifthenelse{\boolean{anon}}{
    \author{\IEEEauthorblockN{Anonymous Authors}
    Submission ID: XX
    }
}
{
    \ifthenelse{\boolean{oneline}}{
        \author{
            \IEEEauthorblockN{Matthieu Bettinger\text{*}\textsuperscript{§}, Etienne Rivière\text{°}, Sonia Ben Mokhtar\text{*}, Anthony Simonet-Boulogne\textsuperscript{\textdagger}}
            \IEEEauthorblockA{\text{*}INSA Lyon, CNRS, Universite Claude Bernard Lyon 1, LIRIS, UMR5205, 69621 Villeurbanne, France\\
            \{given-name\}.\{surname\}@insa-lyon.fr \textsuperscript{§}Corresponding author}
            \IEEEauthorblockA{\text{°}Université Catholique de Louvain, ICTEAM, 1348 Louvain-la-Neuve, Belgium
            \{given-name\}.\{surname\}@uclouvain.be}
            \IEEEauthorblockA{\textsuperscript{\textdagger}iExec Blockchain Tech, 69008 Lyon, France
            \{given-name\}.\{surname\}@iex.ec
            }
            Published in the \emph{2025 25th IEEE International Symposium on Cluster, Cloud and Internet Computing (CCGRID)}\\
            \url{https://doi.org/10.1109/CCGRID64434.2025.00041}
            \vspace{-0.25cm}
            }
    }
    {
        \IEEEauthorblockN{Matthieu Bettinger}
        \IEEEauthorblockA{\textit{LIRIS-DRIM INSA Lyon} \\
        Lyon, France \\
        matthieu.bettinger@insa-lyon.fr}
        \and
        \IEEEauthorblockN{Etienne Rivière}
        \IEEEauthorblockA{\textit{ICTEAM UCLouvain} \\
        Louvain-la-Neuve, Belgium \\
        etienne.riviere@uclouvain.be}
        \and
        \IEEEauthorblockN{Sonia Ben Mokhtar}
        \IEEEauthorblockA{\textit{LIRIS-DRIM CNRS} \\
        Lyon, France \\
        sonia.ben-mokhtar@cnrs.fr}
        \and
        \IEEEauthorblockN{Anthony Simonet-Boulogne}
        \IEEEauthorblockA{\textit{iExec Blockchain Tech} \\
        Lyon, France \\
        anthony.simonet-boulogne@iex.ec}
    }
}

\maketitle

\begin{abstract}
    Current marketplaces rely on search mechanisms with distributed systems but centralized governance, making them vulnerable to attacks, failures, censorship and biases.
    While search mechanisms with more decentralized governance (e.g., DeSearch) have been recently proposed, these are still exposed to \textit{information head-start attacks} (IHS) despite the use of Trusted Execution Environments (TEEs). These attacks allow malicious users to gain a head-start over other users for the discovery of new assets in the market, which give them an unfair advantage in asset acquisition.
    We propose \sysname, a TEE-based provider selection mechanism for distributed search, running in single- or multi-datacenter environments, that is resilient to information head-start attacks. 
    \sysname relies on a Client-TEE collaboration, which enables clients to distinguish between slow providers and malicious ones.
    Performance evaluations in single- and multi-datacenter environments show that, using \sysname, malicious users respectively gain only up to 2\% and 7\% of assets more than without IHS, while they can claim 20\% or more on top of their fair share in the same conditions with DeSearch.    %, as long as there is enough honest provider throughput in the system to sustain the honest user base.
\end{abstract}


\section{Introduction}

Centralized e-commerce platforms (e.g., Amazon or Alibaba) are vulnerable to attacks, failures, and biases~\cite{skim,cuthbertsonFacebookUsersReport2021,swearingenWhenAmazonWeb2018,CensorshipGoogle2024,gargSteemitCensoringUsers2019,glaserHowAppleAmazon2017}. 
% Web3 decentralized marketplaces like OpenSea and Rarible aim to address these issues using blockchain and decentralized storage (e.g., IPFS~\cite{doan2022towards}). 
% However, they often lack integrated search mechanisms or rely on centralized ones, reintroducing vulnerabilities~\cite{stogerDemystifyingWeb3Centralization2023}.
Recent solutions working towards more decentralized governance in search systems~\cite{liBringingDecentralizedSearch2021,keizerDittoDecentralisedSimilarity2023,zichichi_towards_2021} include DeSearch~\cite{liBringingDecentralizedSearch2021}, which uses Trusted Execution Environments (TEEs) to protect against censorship and bias. 
DeSearch relies on users contributing computing resources to run the search protocol. 
These users deploy the DeSearch pipeline on their own TEE-enabled servers or on cloud virtual machines, where TEEs are now widely available (e.g., for Intel SGX, Alibaba, Azure, IBM Cloud, OVH).
These servers crawl the market and maintain an index, on top of which they provide a search service to users.
TEEs at the protocol-level protect the correctness and confidentiality of the search mechanism, 
while the decentralized ownership mitigates attacks from single owners of VMs.

In a context where assets in a marketplace are valuable, scarce, and often both, it is important to ensure that search service providers cannot favor some users and penalize others.
However, the last decade has shown that attacks on marketplaces are commonplace, as soon as there is value to be earned~\cite{eskandariSoKTransparentDishonesty2020,baumSoKMitigationFrontRunning2022}. 
In the context of search, malicious search providers can delay or restrict access to knowledge about new assets for some users, thereby giving a head-start to others, which is comparable to forms of insider trading in financial markets.
We call this behavior an \textit{information head-start} (IHS) attack.
We measure the impact of IHS attacks by the share of new assets discovered by malicious consumers before honest consumers.
We show in this paper that recent work on decentralized search mechanisms for marketplaces is vulnerable to IHS attacks:
despite the use of TEEs, malicious service providers have the power to delay responses to honest consumers, in favor of malicious consumers, which we highlight as a novel attack in the context of TEEs.
Colluding malicious consumers can accentuate the providers' attack by loading honest providers with requests (without a Denial-of-Service attack), increasing those providers' end-to-end latency, so as to steer honest, latency-aware consumers towards malicious providers who will then attack them.

\begin{figure}
  \centering
  \includegraphics[trim={3mm 0 0 0}, scale=0.75]{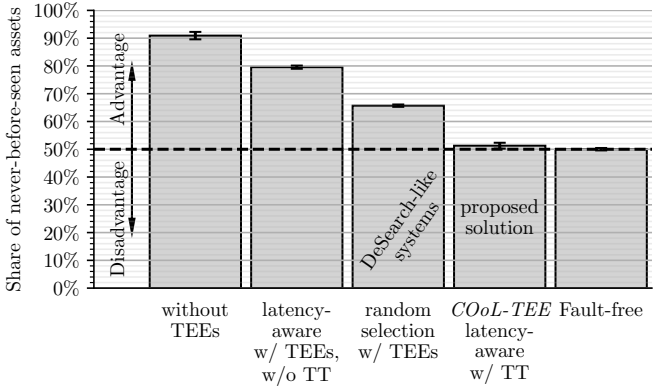}
  \vspace{-0.25cm}
  \caption{Share of never-before-seen assets discovered first by malicious consumers ($\frac{50}{100}$ of total consumers), under different configurations. $\frac{4}{8}$ search providers are malicious and the total system load is 75\%. \emph{The closer the values are to 50\%, the better, indicating lower malicious IHS advantage.}}
  \label{fig:motivation}

\end{figure}

We introduce \sysname, which combines Client-side Optimization of Latencies with provider-side TEEs to protect against IHS attacks. 
\sysname provides a client-side \emph{Provider Selection Module} that selects providers based on past latencies and on trustworthy measurements by TEEs, allowing consumers to avoid high-latency or malicious providers.
%\sonia{ça semble mineur comme ajout, je parlerai du lien avec Desearch plus tard dans le papier, pas forcément dans l'intro}
\Cref{fig:motivation} illustrates the impact of IHS and its mitigation using \sysname in a single datacenter: 
in an unprotected system (1st bar) malicious providers and malicious clients can gain a significant advantage, i.e., discover many new assets before others, due to the absence of integrity and confidentiality guarantees. 
Simply adding TEEs and choosing low-latency providers (2nd bar) is not enough: 
TEEs like Intel SGX do not have access to a Trustworthy Time (TT) source, so malicious providers can manipulate their notion of time and disguise themselves as well-performing providers.
DeSearch's random selection (3rd bar) side-steps being manipulated by malicious providers (by ignoring such misinformation attempts), but is still vulnerable to IHS through delayed responses from randomly selected malicious providers. 
\sysname (4th bar) incorporates a Trusted Time mechanism and mitigates IHS attacks. 
Indeed, \sysname is the closest to the behavior of a fault-free system (5th bar).
While recent work on TEE Trusted Time~\cite{hamidyT3EPracticalSolution2023,fernandezTriadTrustedTimestamps2023a} propose resilience inside a TEE execution against a malicious host adding delays, we contribute a solution that extends resilience to the protocol level, during interactions between remote users and TEEs.

We evaluate \sysname in both cloud and simulated environments (source code available~\cite{coolTEEcode}), comparing its performance against decentralized search and provider selection systems~\cite{liBringingDecentralizedSearch2021,panigrahyAnalysisEvaluationProximitybased2022,stathakopoulouAddingFairnessOrder2021,kelkarThemisFastStrong2023}, under various configurations and attacker strengths.
In particular, we evaluate how much consumers gain or lose in timely market information through IHS, i.e., in opportunities to acquire new assets. 
We present results where up to all providers may be malicious, with 100 consumers and 8 providers in a single datacenter, as well as with 1000 consumers and 48 providers in multiple geo-distributed datacenters. 
We show that, with \sysname, malicious consumers do not get an edge over others in terms of fresher market state information, as long as there is enough honest provider throughput to serve all honest requests, i.e., less than 2\% and 7\% advantage respectively in single- and multi-datacenter experiments, including \Cref{fig:motivation}.
%\sonia{ce serait bien de mettre des chiffres pour illustrer les résultats}

The paper is structured as follows: \Cref{sec:problem_statement} presents the ecosystem and threat models.
\Cref{sec:solution} describes our proposed solution, \sysname. 
\Cref{sec:experimental_protocol,sec:results} cover our experimental protocol and results.
Finally, \Cref{sec:related_work} discusses related work and \Cref{sec:conclusion} concludes the paper.

\section{Model \& problem statement}\label{sec:problem_statement}

We first detail the marketplace search model then formalize our system and attacker models.

\begin{figure}
    \centering
    \includegraphics[scale=1]{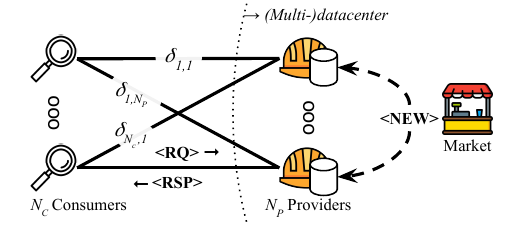}
    \vspace{-0.5cm}
    \caption{\emph{Marketplace search model} Consumers send requests to Providers about the market's state; Providers respond with a list of assets they learned from the Market.}
    \label{fig:system_model}
    \vspace{-0.25cm}
\end{figure}

\subsection{Marketplace search model}
%\sonia{peut être appeler cette section: System Model. Il faudrait clairement ajouter les informations suivantes:
%\begin{itemize}
%    \item On suppose que les indexes de tous les providers sont synchrnisés par un protocole sécurisé et fiable. Les attaques sur ce protocole sont out of scope. 
%    \item 
%\end{itemize}
%}
The system consists of three types of entities, illustrated in Figure~\ref{fig:system_model}: 
(i)~the \emph{market} (right) is the (distributed) storage system where assets are registered. 
(ii)~\emph{Consumers} (left) query the search mechanism;
and (iii)~\emph{Providers} (middle) provide this search mechanism, updating an index of the market state to respond to consumers' requests.

All consumers aim to be the first to discover new assets that appear on the marketplace. 
A subset of consumers tries to improve its asset-discovery chances by colluding with a subset of providers, e.g., in exchange for bribes.
Those subsets of consumers and providers are considered malicious.

We call assets not yet discovered by any consumer \emph{Never-Before-Seen assets} (NBSa). 
When they are first discovered by a consumer, they become a \emph{discovered NBSa} (dNBSa) by this consumer. 
We call attacks impacting a consumer's share of dNBSa \emph{information head-start} (IHS) attacks.
As shown in \Cref{fig:system_model}, we consider $N_{C}$ consumers and $N_{P}$ providers on a network. 
Providers are servers hosted in a (multi-)datacenter environment on TEE-enabled machines,
while consumers are clients either on the same cloud infrastructure or on the Internet.
The network delay between consumer $i$ and provider $j$ is $\delta_{i,j}$.
Two protocols run in parallel: the market and the search mechanism. 
New assets are broadcast to all providers at rate $\lambda_{a}$ as $\mathbf{NEW}$ messages. 
Consumers send requests $\mathbf{RQ}$ to providers at rate $\lambda_{r}$. 
Providers handle requests with a FIFO policy, with a queue delay $\delta_{q}$ and a fixed response time $D$.
Responses $\mathbf{RSP}$ contain assets matching the request's keywords.
We assume consumers only use this search mechanism to discover assets, i.e., they do not crawl the market themselves, due to computational constraints.
The index served by a provider is assumed to be updated synchronously with respect to others providers, e.g., periodically or, if the market is hosted on a blockchain, when a new block appears.

%\sonia{C'est peut être utile de préciser qu'on suppose que le seul moyen de decouvrir des assets est à travers le search mechanism}
\subsection{Trusted Execution Environments (TEEs) and DeSearch~\cite{liBringingDecentralizedSearch2021}}

Intel Software Guard eXtensions~\cite{costan2016intel} (Intel SGX) is a hardware-based Trusted Execution Environment that provides confidentiality and integrity guarantees for code and data running inside that TEE.
DeSearch~\cite{liBringingDecentralizedSearch2021} is a decentralized search protocol using Intel SGX TEEs against censorship and bias attacks. 
DeSearch follows a 3-step pipeline: crawling assets, indexing them, and serving search requests. 
Our work focuses on the search step. 
% Attacks crawling and indexing are out of scope of this work.
Using TEEs, DeSearch guarantees that the index is verifiably correct and complete.
TEEs also ensure search responses are complete and tamper-proof, while Circuit-ORAM~\cite{wangCircuitORAMTightness2015} hides memory access patterns. 
Due to Circuit-ORAM's memory-access obfuscation, concurrent access on the same index is not possible. 
Therefore, in DeSearch and in our model, TEEs serve requests sequentially.

\subsection{Threat model}

\begin{figure*}
    \centering
    \begin{subfigure}{0.3\textwidth}
        \centering
        \includegraphics[scale=1,page=1]{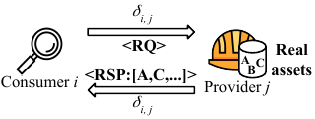}
        \subcaption{Fault-free}
        \label{fig:fault_free}
    \end{subfigure}
    \begin{subfigure}{0.3\textwidth}
        \centering
        \includegraphics[scale=1,page=2]{\figpathattack}
        \subcaption{Content attack}
        \label{fig:content_attack}
    \end{subfigure}
    \begin{subfigure}{0.3\textwidth}
        \centering
        \includegraphics[scale=1,page=3]{\figpathattack}
        \subcaption{Delay attack}
        \label{fig:timing_attack}
    \end{subfigure}
    \vspace{-0cm}
    \caption{Provider-side attack models on marketplace search}
    \label{fig:attack_model}
    \vspace{-0.25cm}
\end{figure*}

Intel SGX augmented with oblivious RAM protects the integrity and confidentiality of the search mechanism.
However, providers that host the TEE retain control over the machine surrounding the TEE enclave: malicious providers can delay responses to honest consumers even in the presence of TEEs.
We identify this attack vector as a lever for IHS attacks.

We assume proportions $c_{M}$ of consumers and $p_{M}$ of providers are malicious. 
We assume honest and malicious actors have similar computing power, proportional to (1-$p_{M}$) and $p_{M}$ for providers, and similarly for consumers with respect to $c_{M}$. 
This computing power corresponds to maximum request emission and handling throughputs respectively for consumers and providers.
All consumers compete over the same stream of assets: preventing honest consumers from discovering NBSa first is a valid strategy for malicious actors.

We now present two provider-side attacks and a consumer-side attack on the marketplace search model and compare them to the fault-free case (\Cref{fig:fault_free}), where all providers serve all consumers with timely and correct results.

\subsubsection{Content attacks} Without TEEs, malicious providers can read requests and send irrelevant, outdated or no responses at all to honest consumers (\Cref{sub@fig:content_attack}). 
This leads to \emph{content attacks}, which comparatively grant more asset-discovery opportunities to malicious consumers.

\subsubsection{Delay attacks} Introducing TEEs protects the protocol's integrity and confidentiality but cannot prevent response delays. 
Indeed, malicious providers, who control the machine around the TEE, can delay responses to honest consumers by $\delta_{\mathit{att}}$ (\Cref{sub@fig:timing_attack}), leading to \emph{delay attacks}. 
Responses are correct but delayed, reducing freshness, i.e., received responses represent an older market state than in the fault-free case.

\subsubsection{Cuckoo attacks} In \emph{cuckoo attacks}, malicious consumers load honest providers with requests, enticing latency-aware honest consumers to switch to malicious providers. 
In turn, these malicious providers can perform a delay attack, resulting in a joint attack which we call a \emph{cuckoo-delay} attack (cuckoo-D). 
For comparison purposes, we also evaluate \emph{cuckoo-content attacks} (cuckoo-C), i.e., the case where providers do not operate TEEs and are so able to perform content attacks.

\subsection{Trusted Time in Trusted Execution Environments}

In TEE-based environments, malicious providers add delays in the protocol for honest consumers to reduce their dNBSa share.
Indeed, without trusted time sources for TEEs, consumers can only trustlessly measure round-trip latencies by themselves.
As a lever for cuckoo-delay attacks, this enables malicious providers to disguise themselves as loaded providers, among the actually-loaded honest providers, victims of the cuckoo-attack's malicious consumer request load.
Already a source of trust inside the potentially malicious host, one would want to also use the TEE to measure delays in the request pipeline and inform consumers.
However, Intel SGX's trusted time primitive~\cite{TrustedTimeMonotonic} is no longer available since 2020 and other time sources are mediated by the untrusted OS~\cite{alderTimeChallengesTemporal2023}.
Recent work~\cite{fernandezTriadTrustedTimestamps2023a,hamidyT3EPracticalSolution2023} tackle restoring a resilient wall clock to Intel SGX TEEs.
Based on DeSearch's per-TEE request throughput in the hundreds per second, we require a fine-grained wall-clock time source in milliseconds that is resilient to TEE-level delay attacks (i.e., a T2 timer in Alder et al.'s TEE Trusted Timer categorization~\cite{alderTimeChallengesTemporal2023}.)
We therefore leverage Triad~\cite{fernandezTriadTrustedTimestamps2023a}, a multi-TEE protocol that avoids disruptions in the TEE's notion of elapsed time through awareness of TEE enclave exits (using AEX-Notify~\cite{constableAEXNotifyThwartingPrecise2023}),
cooperation with TEEs in the same datacenter, and a remote time authority for setup and in cases where all TEEs simultaneously lose their notion of time continuity.
Note that such trusted time mechanisms protect against delay attacks inside the TEE, not at the protocol-level, which is the focus of this paper.
Delay attacks on TEEs themselves~\cite{alderTimeChallengesTemporal2023} operate in a similar way to ours (see \Cref{sub@fig:timing_attack}), the TEE being the consumer and the Trusted Time source the trusted component surrounded by a potential attacker.
With a resilient time source, the TEE can now measure elapsed time for operations occurring inside it and report those measurements to consumers.
\section{COoL | Client-side Optimization of Latencies}\label{sec:solution}

A consumer wants the freshest information about the market, that is, responses that contain the newest relevant assets.
To that end, \sysname enables honest consumers to preferentially select providers that serve responses with lower latencies than others, i.e., that are closer, faster, and honest.
Trustlessly with respect to other consumers and providers hosting the TEEs, \sysname selects providers based on past latencies the consumer personally experienced and on request residence time measurements from TEEs.

To present our solution, we first focus on the timing aspects of the two parallel protocols, as shown in \Cref{fig:system_arch}: market indexing (right) and our proposed provider-selection and search mechanisms (left). 
We then detail the \emph{Provider Selection Module} (PSM) that performs Client-side Optimization of Latencies, leveraging TEE Trusted Time.

\subsection{Protocol delays and timing}\label{subsec:search}

\begin{figure}
    \centering
    \includegraphics[trim={2mm 2mm 0 2mm},scale=1]{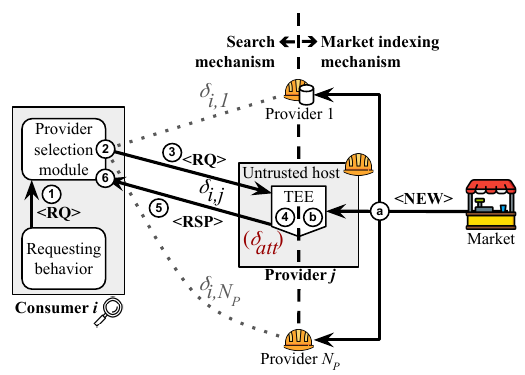}
    \caption{High-level system architecture}
    \vspace{-0.25cm}
    \label{fig:system_arch}
    \vspace{-0cm}
\end{figure}

\Cref{fig:system_arch} shows the behavior of a consumer $i$ using the search mechanism and providers 1 to $N_{P}$: 
(1)~At $t_{\mathit{gen}}$, consumer $i$ generates a request $\mathbf{RQ}$.
(2)~The PSM directs the request to provider $j$ at $t_{\mathit{send}}$ after a delay $\delta_{\mathit{PSM}}$. 
(3)~The request is sent to provider $j$ with a one-way network delay $\delta_{i,j}$. 
(4)~The request waits in the TEE's queue for $\delta_{\mathit{queue}}$ until service starts at $t_{serv}$, with a service time $\delta_{\mathit{serv}}=D$.
The TEE measures the request's waiting time $\delta_{\mathit{queue}}$ and reports it in the response.
(5)~After service, the TEE sends the response $\mathbf{RSP}$, which may be delayed by $\delta_{att}$ if the host is malicious. 
The response is then sent back to consumer $i$ with a delay $\delta_{i,j}$.
(6)~At $t_{\mathit{recv}}$, consumer $i$ receives the response and logs the round-trip latency minus the queue waiting time as $\Delta^{k}_{i,j}$, the $k^{th}$ measurement with respect to provider $j$.

Concurrently, new assets are broadcast to all providers (step~a) and indexed at $t_{index}$ (step~b). 
For an asset $A$ to appear in a response, $t_{index}<t_{\mathit{serv}}$. 
The freshness of a response is $\delta_{\mathit{fresh}}=D+\delta_{i,j}+\delta_{att}$, i.e., the delay from when the TEE starts serving a request to when the consumer receives the response. 
Honest consumers maximize freshness by choosing close, fast, and honest providers.
As it is difficult to trustlessly measure the individual delays that make up $\delta_{\mathit{fresh}}$, notably the return-trip outside the TEE $\delta_{i,j}+\delta_{att}$ (consumer and TEE clock phases are not assumed to be synchronized), honest consumers rely with \sysname on minimizing round-trip latencies substracted by TEE-measured trustworthy delays.
To appropriately substract the TEE-measured queue waiting time from the consumer-measured round-trip latency, as part of the communication setup with each TEE, e.g., after TEE attestation, a consumer measures the drift rate with the TEE's time source.
To that end, subprotocols of clock synchronization protocols like NTPsec\cite{raymondNtpsecSecureHardened2016a} can be used, e.g., in that case, the clock filter and discipline algorithms~\cite{ClockDisciplineAlgorithm}.

%\subsection{Leveraging TEEs and Trusted Time}

%Consumers cannot trust providers hosting search services. 
%However, TEEs within the potentially malicious host can relay information about the service to consumers, as long as that information cannot be manipulated by the rest of the malicious system.
%Recall that consumers want to minimize $\delta_{fresh}$, but can only trustlessly measure the whole round-trip latency $D$ by themselves.
%Part of that round-trip latency, $\delta_{queue}+\delta_{serv}$, happens within the TEE.
%Using Triad~\cite{fernandezTriadTrustedTimestamps2023a} to prevent time manipulations by the host, in \sysname, for each consumer request, the TEE measures the queuing delay $\delta_{queue}$ and sends it as part of the response.
%Consumers then substract $\delta_{queue}$ from $D$ to better estimate $\delta_{fresh}$.

\subsection{COoL provider selection}\label{subsec:cool}

A key mechanism in \sysname is the Provider Selection Module (PSM). 
Consumers want the freshest information about the market's state. 
%To that end, the PSM prefers providers with low latencies and so, among providers with similar computing power ($\delta_{\mathit{serv}}$) and network conditions ($\delta_{i,j}$), avoids malicious providers with $\delta_{\mathit{att}}>0$. 
To minimize average response latencies, each consumer $i$'s PSM uses \emph{Client-side Optimization of Latencies} (COoL) by aggregating past round-trip latencies and TEE delay measurements as $\Delta^{k}_{i,j}$ and then adjusting the probability of selecting each provider. 
Consumer $i$ then directs a proportion $r_{i,j}$ of their total request throughput $\lambda_{r}^{i}$ towards each provider $j$. 
%As the incoming request load increases at a provider, so does the queuing delay at a provider's TEE, increasing round-trip latency. 
%Hence, there exists a valuation of all $r_{i,j}$ that minimizes average round-trip latency.
In its implementation~\cite{coolTEEcode}, \sysname uses a sliding-window average of the $s$ latest latencies from each provider to compute the average latencies. 

\begin{algorithm}
    \KwIn{$\{ r_{j}^{in} \}$ set of initial provider selection ratios; $\{\Delta_{j}\}$ set of per-provider observed latencies sets}
    \KwData{$x$ exploration coefficient; $K_{p}$, $K_{d}$ PD-controller coefficients; $s$ sliding window size; $c$ cluster threshold; $a$ attrition ratio}
    \KwOut{$\{ r_{j}^{out} \}$ set of updated provider selection ratios}
    
    \Begin{
        \ForEach{$j \in \llbracket 1;N_{P}\rrbracket$}{
            $\Delta_{j}^{\mathit{avg}} \gets \frac{1}{s}\sum_{k=1}^{s}\Delta_{j}^{k}$\;
            $\Delta_{j}^{\mathit{prev},\mathit{avg}} \gets \frac{1}{s}\sum_{k=s+1}^{2s}\Delta_{j}^{k}$\;
        }
        \tcc{Cluster best providers in $B_{prov}$ and worst providers in $W_{prov}$}
        $\Delta_{best}^{avg} \gets \min_{j \in \llbracket 1;N_{P}\rrbracket}(\Delta_{j}^{avg})$\;
        $B_{prov} \gets \{ j \in \llbracket 1;N_{P}\rrbracket \mid \Delta_{j}^{avg} \leq \Delta_{best}^{avg} + c \}$\;
        $W_{prov} \gets \{ j \in \llbracket 1;N_{P}\rrbracket \mid \Delta_{j}^{avg} > \Delta_{best}^{avg} + c \}$\;
        $\Delta^{\mathit{avg}}_{target} \gets \frac{1}{|B_{prov}|}\sum_{j\in B_{prov}}\Delta_{j}^{\mathit{avg}}$\;
        \tcc{PD-control best providers}
        \ForEach{$j \in B_{prov}$}{
            $e_{j} \gets \frac{\Delta^{\mathit{avg}}_{target} - \Delta_{j}^{\mathit{avg}}}{\Delta^{\mathit{avg}}_{target}}$\;
            $d_{j} \gets \Delta_{j}^{\mathit{avg}} - \Delta_{j}^{prev,\mathit{avg}}$\;
            $r_{j}^{\mathit{tmp}} \gets clamp(r_{j}^{in} + K_{p}e_{j} + K_{d}d_{j}$, 0, 1)\;
        }
        $S_{best} \gets \sum_{j\in B_{prov}}r_{j}^{\mathit{tmp}}$\;
        \tcc{Lessen use of worst providers}
        \ForEach{$j \in W_{prov}$}{
            $r_{j}^{norm} \gets (1-a)r_{j}^{in}$ \;
        }
        $S_{worst} \gets \sum_{j\in W_{prov}}r_{j}^{\mathit{tmp}}$\;
        \ForEach{$j \in B_{prov}$}{
            $r_{j}^{\mathit{norm}} \gets (1-S_{worst})r_{j}^{\mathit{tmp}}/S_{best}$\;
        }
        \ForEach{$j \in \llbracket 1;N_{P}\rrbracket$}{
            $r_{j}^{out} \gets (1-x)r_{j}^{\mathit{norm}} + x\frac{1}{N_{P}}$\;
        }
        \Return{$\{ r_{j}^{out} \}$}
    }
    \caption{PSM.UpdateSelectionRatios}
    \label{algo:update_ratios}
\end{algorithm}

The PSM operates in two phases: selecting a provider for each request and updating selection ratios.
In \Cref{fig:system_arch}, at step 2 of the search mechanism, the PSM selects a provider $j$ with probability $r_{j}$.
Periodically, the PSM uses \Cref{algo:update_ratios} to update $\{r_{j}\}$ based on observed latencies $\{\Delta_{j}\}$. 
In lines 3 and 4, average per-provider latency in the current and previous windows are computed respectively as $\Delta_{j}^{\mathit{avg}}$ and $\Delta_{j}^{\mathit{prev,\mathit{avg}}}$.
Based on the provider with the lowest average latency $\Delta^{avg}_{best}$ (line 5), providers are divided in two groups: the ``best providers'', whose average latency is at most $\Delta^{avg}_{best}+c$, in $B_{prov}$; while the rest constitute the ``worst providers'', in $W_{prov}$.
Lines 9 to 12 implement a Proportional-Derivative (PD) controller on selection ratios for the best providers. 
The PD-controller's setpoint is the current average latency $\Delta^{\mathit{avg}}_{target}$ among providers in $B_{prov}$ (line 8).
The provider-specific relative error $e_{j}$ to the target (line 10) as well as its derivative $d_{j}$ (line 11) are computed to adjust the selection. 
The result of the Proportional Derivative control expression $r_{j}^{in} + K_{p}e_{j} + K_{d}d_{j}$ is then mapped to its closest value in the $[0,1]$ interval (line 12).
If a provider's latency $\Delta_{j}^{avg}$ is higher than the average $\Delta_{target}^{avg}$, $r_{j}$ decreases; if it is lower, $r_{j}$ increases. 
Meanwhile, the worst providers lose a portion $a$ of their selection ratios (line 15) and this loss is transferred to the best providers during a normalization step (line 18). 
As a last common step for all providers, selection ratios are mixed with a uniform probability (line 20): the PSM allocates a proportion $x$ of requests to explore the latency Quality-of-Service at available providers.
Indeed, we remark that this client-side provider selection can be described as a Multi-Armed Bandit problem~\cite{slivkins2019introduction}: a consumer sends a request to a chosen provider $j$ (i.e., pulls lever $j$), then experiences a round-trip latency and a TEE-reported queue waiting-time (i.e., the reward).
The consumer's objective is to maximize its reward: select providers in order to minimize average latencies. 
As is common in Multi-armed bandit settings~\cite{slivkins2019introduction}, the selection policy should allocate a small fraction of requests to explore the latency-performance of each provider, to avoid exploiting sub-optimal providers asymptotically.

\Cref{algo:update_ratios} requires calibration of the coefficients $K_{p}$ and $K_{d}$, as well as of the exploration coefficient $x$, so as to offer theoretical guarantees.
This can be online, during the system's operation, with parameter-tuning algorithms~\cite{fiducioso2019safe,xu2023config}. 
Algorithms from the Multi-Armed Bandit literature that tolerate evolving reward distributions, like Exp3~\cite{auer2002nonstochastic}, may also be used. 
As our focus in this paper is not on convergence itself, but on the asymptotic behavior of the system where selection ratios have converged, we tune parameters manually. 
We obtain a system where the selection ratios converge to the valuations yielding minimal latencies in micro-benchmark scenarios, i.e., in a set of stereotypical scenarios (e.g., high- versus low-latency providers; high- versus low-throughput providers), and converge with random topologies drawn from the dataset which will be presented in the next section.

\section{Experimental protocol}\label{sec:experimental_protocol}

This section presents how we evaluate \sysname. 

\subsubsection{Metrics}
We use discovered Never-Before-Seen assets (dNBSa) as well as the system's latency and throughput as metrics to evaluate \sysname with respect to related work.
The main metric is the number of dNBSa discovered by malicious consumers. 
If the proportion of dNBSa by malicious consumers matches their proportion of sent requests among all the whole consumer population, there is no IHS advantage. 
Higher or lower proportions indicate IHS advantage or disadvantage, respectively. 
Secondary metrics are latency and peak throughput for consumer requests and provider responses, impacting response freshness and Quality-of-Service (QoS).

\subsubsection{Deployment and simulation setups}
We implement \sysname in C++, using DeSearch~\cite{liBringingDecentralizedSearch2021} as the backend. 
We deploy the system on Ubuntu 20.04 Azure VMs, where providers run on a 4-core SGX-enabled VM (\emph{DC4s\_v2}), and consumers on a 16-core non-SGX VM (\emph{B16als\_v2}). 
Both machines are in the same datacenter, with an average round-trip latency of 1.1ms (0.5ms of standard deviation in 100 pings) between them.
SGX's AEX-Notify~\cite{constableAEXNotifyThwartingPrecise2023}, required by Triad's trusted time mechanism~\cite{fernandezTriadTrustedTimestamps2023a}, is only available for Linux kernels 6.2+, but DeSearch~\cite{liBringingDecentralizedSearch2021} has dependencies with Ubuntu 20.x. 
Therefore, we evaluate the impact of Trusted Time on a separate 32-core SGX2-enabled server and simulate DeSearch in that case as a black box with the same interface.
We use these deployments to measure overheads incurred by our protocol in a distributed setting and to calibrate parameters of a simulated version of the system.
Indeed, due to the high number of configurations to evaluate, we also implement a simulated version of \sysname and of baselines using Omnet++~\cite{vargaOverviewOMNeTSimulation2008}:
simulation experiment design points presented in this paper represent around 40k protocol runs and 1TB of requests' lifecycle traces.
The code and analysis scripts are publicly available~\cite{coolTEEcode}.
Absent in the original Triad paper~\cite{fernandezTriadTrustedTimestamps2023a}, we also contribute a public implementation of Triad's trusted time protocol, that we integrate to COoL-TEE~\cite{coolTEEcode}.

\subsubsection{Network topologies}
We consider two network topologies: (1) ``same-datacenter'' topology, where latencies are the same between all actors ($\delta_{\mathit{net}}=0$) and (2) ``multi-datacenter'' topology, where latencies between consumers and providers are heterogeneous and assigned using the Cloud-Edge latency dataset~\cite{charyyevLatencyComparisonCloud2020}.
This dataset contains the median roundtrip latencies between 69 datacenters from major cloud service providers (CSPs) and 8456 probes from RIPE Atlas~\cite{RIPEAtlasRIPE}, an Internet connectivity monitoring service, all positioned across the globe.
In simulation-based experiments, our search providers are assigned to datacenter nodes in that dataset, while consumers are a subset of RIPE Atlas probes.
To still take advantage of the datacenter geodistribution by CSPs among a smaller subset of datacenters, without loss of generality, we use IBM datacenters from the dataset (22 datacenters, the most numerous compared to other represented cloud service providers).
We position 2 provider nodes per datacenter, except in South America (4 nodes) and in Australia (3 nodes) to account for the low number of existing IBM datacenters there (resp. 1 and 2).
Additionally, we sample equal numbers of consumer nodes whose closest datacenter is in the same geographic region (i.e., 200 consumers each, in North America, South America, Europe, Asia, and Oceania.)
Using that dataset, the latency distribution for users to their closest datacenter depends on the geographic region: the median latency is around 15ms to 30ms for all regions except South America (40ms).
Depending on the geographic region, 80 to 95\% of users access their closest datacenter with less than 50ms round-trip latency (60\% for South America,) while 98\% of users have less than 150ms round-trip latency.

\begin{figure*}
    \centering
    \begin{subfigure}{0.49\textwidth}
    \centering
    \includegraphics[trim={2mm 2mm 0 2mm}, scale=0.85]{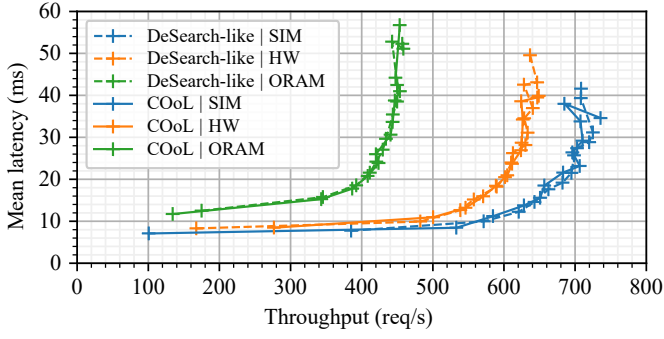}
    \vspace{-0.5cm}
    \caption{Impact of SGX and ORAM~\cite{wangCircuitORAMTightness2015} on DeSearch and \sysname.}
    \label{fig:ff-lat-through-sgx}
\end{subfigure}
\begin{subfigure}{0.49\textwidth}
    \centering
    \includegraphics[trim={2mm 2mm 0 2mm}, scale=0.85]{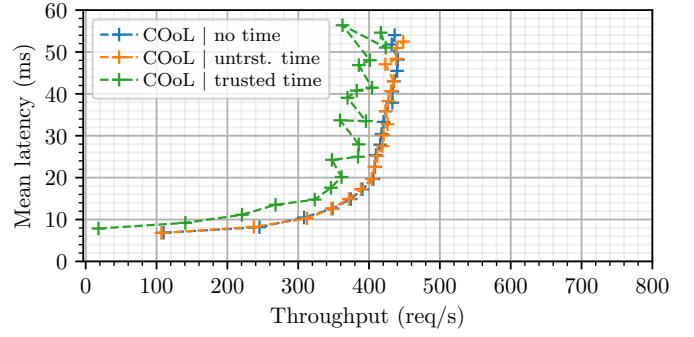}
    \vspace{-0.5cm}
    \caption{Impact of the time source type on \sysname provider selection.}
    \label{fig:ff-lat-through-TT}
\end{subfigure}
\caption{Mean latency versus throughput depending on enabled security and time mechanisms, with resp. $N_{P}=4$ and 3.}
\vspace{-0.25cm}
\label{fig:ff-lat-through}
\end{figure*}

\subsubsection{Simulation parameters}
Simulations involve $N_{C}=100$ consumers and $N_{P}=8$ providers for same-datacenter experiments and $N_{C}=1000$ consumers and $N_{P}=48$ providers for multi-datacenter experiments. 
Each provider handles up to 160 requests per second (6.25~ms/request), a throughput obtained empirically with DeSearch.
Consumers send requests at rates $\lambda_{r}$ so as to ensure a total average system load $\rho=75\%$ (unless otherwise specified). 
On the market-side, assets arrive at an average $\lambda_{a}=100$ assets per second.
Regarding event distributions, figures present results for consumer requests emission and asset arrivals following Poisson processes, with respective rates $\lambda_{r}$ and $\lambda_{a}$.
We do not illustrate results in the paper for other distributions due to lack of space: with periodic events (whether requests or assets), results are similar on average to those presented, but with higher standard deviations.

\subsubsection{Provider selection policies}
Consumers select providers using ``DeSearch-like'' random selection or \sysname's ``COoL'' selection optimizing for latency. 
We assess ``COoL'' selection both without and with Trusted Time (TT).
We also separately evaluate COoL selection augmented with state-of-the-art Power-of-Two (PoT) load balancing~\cite{mitzenmacherPowerTwoChoices2001,panigrahyAnalysisEvaluationProximitybased2022}, called ``COoL-PoT''. 
In COoL-PoT, consumers select two providers, and the one with the shorter queue handles the request. 
Finally, we consider multiprovider selection policies where honest consumers send requests to $k$ providers to probabilistically avoid sending requests only to malicious providers.

\subsubsection{Attack configurations}
The malicious providers' proportion $p_{M}$ varies between 0 and 1. 
Meanwhile, in illustrated results, malicious consumers' proportion $c_{M}$ in the population is set to 50\% for more intuitive result description, but we discuss the influence of $c_{M}$ on IHS.
We implement provider- and consumer-side attack strategies defined in \Cref{sec:problem_statement}. 
Delay attacks use a delay $\delta_{\mathit{att}}=50ms$.
Additionally, for the specific case of selection policies using Power-of-Two, we consider the attack vector that opens using that mechanism, which we call the ``queue attack''.
In a queue attack, malicious providers keep a secondary queue outside the TEE, so that the TEE's queue always remains at a length $L\leq1$ (considering a request being handled remains at the front of the queue until it is served).
This means that from the point of view of honest providers, malicious providers more often have shorter queues than them and, consequently, should yield service of a request to them more often.

\section{Results}\label{sec:results}

We evaluate \sysname's performance in terms of IHS advantages and latency-throughput to answer:
\begin{itemize}
    \item[]\emph{RQ1}: What is \sysname's performance in a fault-free case?
    \item[]\emph{RQ2}: What is its performance in an adversarial same-datacenter setting?
    \item[]\emph{RQ3}: What is its performance in an adversarial multi-datacenter setting?
\end{itemize}
%We answer research questions RQ1 to RQ3 respectively in \Cref{ssec:res-FFPL,ssec:res-sameDC,ssec:res-malPL}.

\subsection{Fault-free setup}\label{ssec:res-FFPL}

To address \emph{RQ1}, we first evaluate the impact of solution building blocks in a fault-free setting deployed on the cloud.
%We compare building block overheads on the same-datacenter topology.
%\subsubsection{Building block overheads}\label{ssec:res-bbo}
\subsubsection{Provider selection and security mechanism overheads}
\Cref{fig:ff-lat-through-sgx} shows overheads for provider selection policies and security features in a same-datacenter environment. 
Latency is measured from request generation to response reception, including provider selection time. 
The choice of selection policy has neglible impact on maximum throughput and average latency. 
Compared to an execution in SGX simulation mode (``SIM'' in \Cref{fig:ff-lat-through-sgx}), SGX hardware mode without Circuit-ORAM (``HW'') has a low latency overhead ($<2ms$) and reduces throughput by around 11\%. 
Compared to standalone SGX hardware mode, Circuit-ORAM increases latency by 4 to 10ms and reduces throughput by 30\%.

\subsubsection{Time source overheads} Following the provider selection and SGX overhead analysis, we evaluate in \Cref{fig:ff-lat-through-TT} the impact of the time source type on \sysname's performance.
Without measuring time and when using the untrusted OS time, i.e., without deploying the Triad~\cite{fernandezTriadTrustedTimestamps2023a} Trusted Time mechanism alongside COoL-TEE, we obtain a total peak throughput of around 440k requests per second for $N_{P}=3$ providers. 
Using Trusted Time, the throughput is reduced by around 10 to 15\%, while the latency increases by 2 to 4ms.

\subsection{Adversarial same-datacenter setup}\label{ssec:res-sameDC}

\begin{figure*}
    \centering
    \begin{subfigure}{0.49\textwidth}
        \centering
        \includegraphics[trim={2mm 2mm 0 0mm}, scale=0.78]{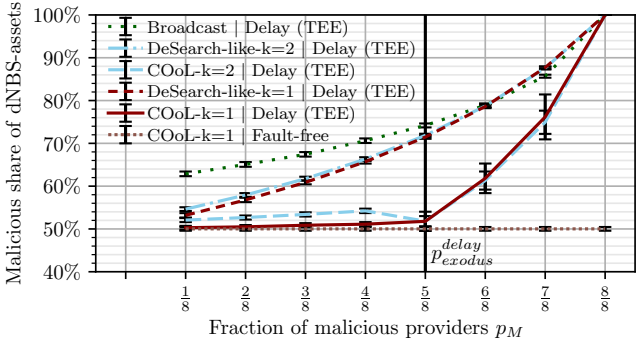}
        \subcaption{Impact of TEE \emph{delay attacks} on single-/multi-provider selection}
        \label{fig:byz-sameDC-cs-rw-multi}
    \end{subfigure}
    \begin{subfigure}{0.49\textwidth}
        \centering
        \includegraphics[trim={2mm 2mm 0 0mm}, scale=0.78]{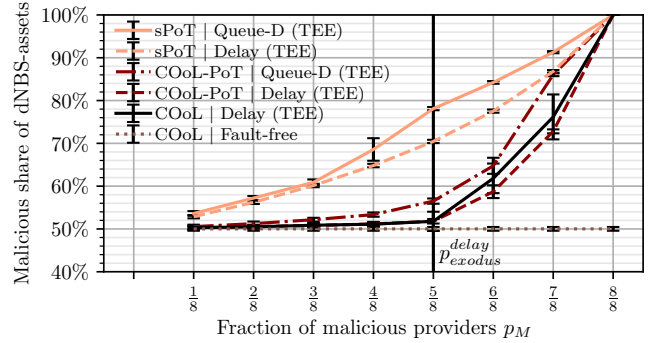}
        \subcaption{Impact of TEE-based attacks on sPoT~\cite{panigrahyAnalysisEvaluationProximitybased2022} and COoL selections}
        \label{fig:byz-sameDC-cs-rw-spot}
    \end{subfigure}
    \caption{Information head-start case-study with respect to competitors in a \emph{delay-attacked} TEE environment | Shares of discovered NBSa across 100 runs, in a malicious same-datacenter environment with 8 providers. \emph{The closer the values are to 50\%, the better, indicating lower malicious IHS advantage.}}
    \label{fig:byz-sameDC}
    \vspace{-0.25cm}
\end{figure*}

We now simulate attacks in a same-datacenter setup with varying proportions of malicious providers $p_{M}$, evaluating the impact of provider selection policies, of load balancing policies, of attack strategies, and of scaling out providers.
We first evaluate systems without Trusted Time in \Cref{ssec:res-multi,ssec:res-spot,ssec:cs3,ssec:cs4}, then with Trusted Time in \Cref{ssec:t4ct}.
\subsubsection{Multiprovider selection}\label{ssec:res-multi}
We compare multiprovider selection with De\-Search-like or COoL policies, while varying $k\in \{1,2,8\}$, the number of selected providers per request. 
In the 8-provider setup, $k=8$ corresponds to the broadcast policy used in related work~\cite{stathakopoulouAddingFairnessOrder2021,kelkarThemisFastStrong2023}.
\Cref{fig:byz-sameDC-cs-rw-multi} shows the information head-start (IHS) advantage for malicious consumers, measured by their share of dNBSa, under \emph{delay attacks} and in the \emph{fault-free} case. 
Higher $k$ values increase the malicious advantage. 
Single-provider COoL selection ($k=1$) performs best: as long as $p_{M}\leq p_{\mathit{exodus}}^{\mathit{delay}}$, the IHS advantage remains within 2\% of the fault-free case at 50\%.
$p_{\mathit{exodus}}^{\mathit{delay}}$ is the proportion of malicious providers beyond which honest provider throughput is insufficient to serve all honest requests under \emph{provider-side delay attacks}. 
When $p_{M}> p_{\mathit{exodus}}^{\mathit{delay}}$, i.e., with more than 5 malicious providers, honest requests are increasingly served by malicious providers, increasing in turn malicious IHS. 
In contrast, DeSearch-like selection with any $k$ and COoL multiprovider selection with $k>2$ both start with higher IHS advantage and show a steady increase with $p_{M}$.

\subsubsection{Provider-side load balancing}\label{ssec:res-spot}
In \Cref{fig:byz-sameDC-cs-rw-spot}, we compare spatial Power-of-two~\cite{panigrahyAnalysisEvaluationProximitybased2022} (sPoT) to our proposed solution in a \emph{delay-attacked} TEE environment. 
We then also evaluate the impact of the new attack introduced by provider-side load balancing: the \emph{queue attack} coupled with \emph{delay attacks}.

First, with only \emph{delay attacks} and with standalone sPoT~\cite{panigrahyAnalysisEvaluationProximitybased2022}, the share of dNBSa by malicious consumers increases steadily with $p_{M}$.
In sPoT, consumers rely on a network distance measurement (e.g., a ping) to determine their two closest providers.
In an adversarial same-datacenter setup, where all providers are roughly equidistant to a given consumer, and can lie about one-shot latency measurements, sPoT amounts to single-provider random selection: ``DeSearch-like-k=1'' in \Cref{fig:byz-sameDC-cs-rw-multi} and ``sPoT'' in \Cref{fig:byz-sameDC-cs-rw-spot} have similar shapes.  
Meanwhile, with consumer-side COoL selection, as well as with its $k=2$ variant augmented with sPoT at the provider-side, the malicious dNBSa share remains close to the fault-free case as long as $p_{M}\leq p_{\mathit{exodus}}^{\mathit{delay}}$.

However, when we consider the increased attack surface introduced with provider-side load balancing, i.e., 
when malicious providers leverage \emph{queue attacks} on top of \emph{delay} attacks, solutions based on this provider-side balancing suffer more from IHS.
Standalone ``COoL'' selection is client-side only, so this attack vector does not exist.

\subsubsection{COoL selection under fire}\label{ssec:cs3}

\begin{figure*}
    \centering
    \begin{subfigure}{0.49\textwidth}
        \centering
        \includegraphics[trim={2mm 2mm 0 0mm}, scale=0.78]{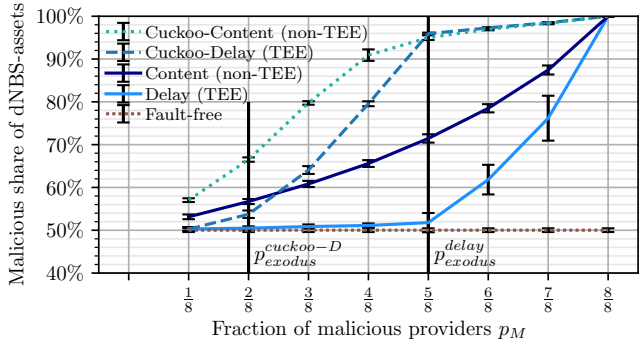}
        \subcaption{IHS attacks on COoL provider selection with 8 total providers}
        \label{fig:byz-sameDC-cs-8sp}
    \end{subfigure}
    \begin{subfigure}{0.49\textwidth}
        \centering
        \includegraphics[trim={2mm 2mm 0 0mm}, scale=0.78]{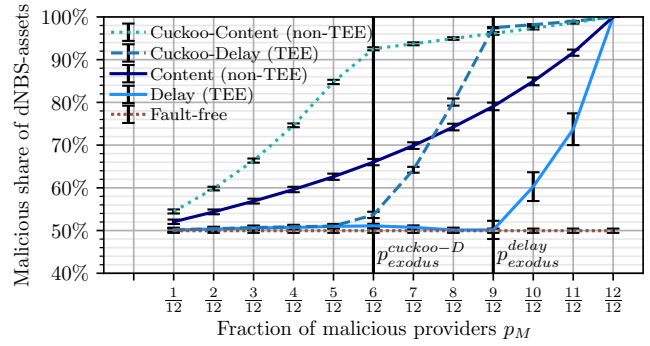}
        \subcaption{IHS attacks on COoL provider selection, scaled out by 4 providers}
        \label{fig:byz-sameDC-cs-12sp}
    \end{subfigure}
    \caption{Information head-start case-study with respect to attack scenarios in (non-)TEE environments | Shares of discovered NBSa across 100 runs, in a malicious same-datacenter environment, with either 8 or 12 providers (resp. \Cref{fig:byz-sameDC-cs-8sp,fig:byz-sameDC-cs-12sp}).}
    \label{fig:byz-sameDC-cs-att}
    \vspace{-0.25cm}
\end{figure*}

\Cref{fig:byz-sameDC-cs-8sp} focuses on IHS advantage using single-provider COoL selection, when subject to different provider- and consumer-side attacks. 
\emph{Content-} and \emph{delay-based attacks} show different behaviors: while \emph{(cuckoo-)content attacks} have the share of dNBSa grow steadily with $p_{M}$, \emph{(cuckoo-)delay attacks} have this share close to the fault-free case, so long as $p_{M}$ is less than the attack-specific threshold $p_{\mathit{exodus}}$.
After their $p_{\mathit{exodus}}$, the IHS advantage rises sharply for each \emph{delay-based attack}.
Note that $p_{\mathit{exodus}}^{\mathit{cuckoo-D}}<p_{\mathit{exodus}}^{\mathit{delay}}$. 
Without cuckoo attacks, malicious consumers aim to minimize latency, driving them towards malicious providers, which honest consumers avoid. 
In \emph{cuckoo-delay attacks}, malicious consumers prioritize loading honest providers with requests over latency minimization. 
Thus, honest consumers shift towards malicious providers at a lower $p_{M}$, i.e., $p_{\mathit{exodus}}^{\mathit{cuckoo-D}}<p_{\mathit{exodus}}^{\mathit{delay}}$. 
In \Cref{fig:byz-sameDC-cs-8sp}, the impact of \emph{cuckoo-delay attacks} at a given $p_{M}$ is similar to a \emph{provider-side-only delay attack} with three additional malicious providers. 
Hence, a lower $p_{M}$ achieves the same IHS advantage when malicious consumers actively participate in attacks.

\subsubsection{Provider scale-out}\label{ssec:cs4}

\Cref{fig:byz-sameDC-cs-12sp} evaluates the impact of scaling out providers in response to high loads: 4 additional providers are launched compared to \Cref{fig:byz-sameDC-cs-8sp}. 
The incoming consumer load remains the same, reducing the total system load $\rho$ from 75\% (8 providers) to 50\% (12 providers). 
Lowering $\rho$ raises the minimum thresholds $p_{\mathit{exodus}}^{\mathit{cuckoo-D}}$ and $p_{\mathit{exodus}}^{\mathit{delay}}$ for impactful attacks: $p_{M}>25\%$ at $\rho=75\%$, while $p_{M}>50\%$ at $\rho=50\%$. 
$p_{\mathit{exodus}}^{\mathit{cuckoo-D}}$ is tied to $\rho$, marking when honest consumers cannot handle the total request load: $p_{\mathit{exodus}}^{\mathit{cuckoo-D}}= 1-\rho$. $p_{\mathit{exodus}}^{\mathit{delay}}$ depends on $\rho$ and $c_{M}$, as only honest consumers load honest providers with requests in provider-side-only delay attacks: $p_{\mathit{exodus}}^{\mathit{delay}}= 1-(1-c_{M})\rho$. 
Without TEEs, under (cuckoo-)content attacks, malicious dNBSa shares are not sensitive to $\rho$: malicious IHS advantage depends on $p_{M}$ for provider-side content attacks, and on $p_{M}$ and $c_{M}$ for cuckoo-content attacks. 
The slope change in \Cref{fig:byz-sameDC-cs-8sp,fig:byz-sameDC-cs-12sp} occurs when the proportion of malicious providers equals the proportion of honest consumers: $p_{\mathit{exodus}}^{\mathit{cuckoo-C-satur}}= 1-c_{M}$.

\subsubsection{Trusted Time for \sysname}\label{ssec:t4ct}

\begin{figure*}
    \centering
    \begin{subfigure}{0.49\textwidth}
        \centering
        \includegraphics[trim={2mm 2mm 0 0mm}, scale=0.78]{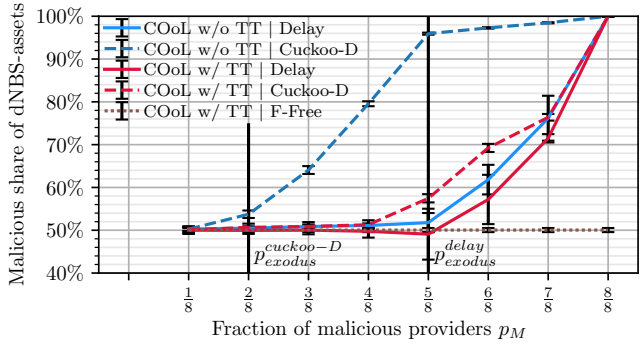}
        \caption{8 providers in a single \emph{(cuckoo-)delay-attacked} datacenter.}
        \label{fig:byz-sameDC-cs-att-TT}
    \end{subfigure}
    \begin{subfigure}{0.49\textwidth}
    \centering
        \includegraphics[trim={2mm 2mm 0 0mm}, scale=0.78]{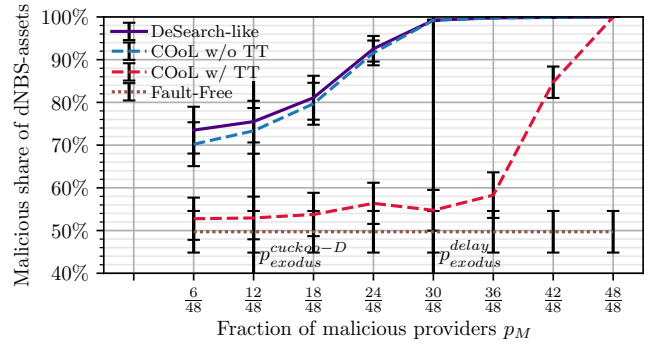}
        \caption{48 providers in multiple \emph{cuckoo-delay-attacked} datacenters.}
        \label{fig:byz-multiDC-cs-att}
    \end{subfigure}
    \caption{Information head-start case-study with respect to attacks on \sysname with and without Trusted Time (TT) | Shares of discovered NBSa across 100 runs, in a malicious same- or multi-datacenter environment (resp. \Cref{fig:byz-sameDC-cs-att-TT} and \Cref{fig:byz-multiDC-cs-att}).}
    \label{fig:byz-TT}
    \vspace{-0.25cm}
\end{figure*}

Finally, we evaluate how measuring trustworthy delays and relaying them to consumers impacts IHS in \Cref{fig:byz-sameDC-cs-att-TT}.
We show IHS for \sysname with and without Trusted Time (``w/ TT'' and ``w/o TT'', respectively), under provider-side delay attacks as well as cuckoo-delay attacks.
``COoL w/o TT'' curves are the same as in \Cref{fig:byz-sameDC-cs-8sp}.
Adding trusted time to \sysname, resilience against IHS in provider-side delay attacks slightly improves, by up to 4\% less malicious dNBSa share.
Against the more powerful cuckoo-delay attack, resilience is greatly improved with trusted time: ``COoL w/ TT'' under cuckoo-delay attacks behaves similarly to ``COoL w/o TT'' under the weaker provider-side delay attacks.
Similarly to previous figures, the curve's change in trend happens around $p_{\mathit{exodus}}^{\mathit{delay}}$.

\subsubsection{Case-studies summary}
To summarize and answer \emph{RQ2}, in the same-datacenter setup, with COoL provider selection, \emph{delay-attacked} honest consumers discover NBSa at similar rates as malicious consumers, as long as the honest providers' maximum throughput is sufficient to serve either: all requests without trusted time or all \emph{honest} requests with trusted time.
In practice, this can be ensured using load-aware elasticity mechanisms~\cite{HorizontalPodAutoscaling}: the fourth case-study in \Cref{ssec:cs4} and \Cref{fig:byz-sameDC-cs-att} show how reducing the total system load $\rho$ increases the necessary malicious provider proportion $p_{M}$ for impactful attacks.
Meanwhile, without TEEs, or without latency-aware provider selection, the discovery of dNBSa by malicious consumers increase nearly linearly with respect to the proportion of malicious providers $p_{M}$.
Experiments were also run either with periodic asset arrivals, or periodic inter-request emission delays, but we do not illustrate them due to lack of space: conclusions drawn above also hold in those cases, albeit with higher standard deviations. 

\subsection{Adversarial multi-datacenter setup}\label{ssec:res-multiDC}

To answer \emph{RQ3}, we now analyze IHS in a multi-datacenter setting in \Cref{fig:byz-multiDC-cs-att}, under a cuckoo-delay attack.
The total system load is $\rho=75\%$ like in \Cref{fig:byz-sameDC-cs-att-TT}.
We compare both figures in the following analysis.
For a fairer comparison to DeSearch's random selection in a context with geo-distributed datacenters,
DeSearch consumers in this experiment only randomly select among their personal best-performing providers, i.e., among those in \Cref{algo:update_ratios}'s $B_{prov}$ (line 6).
We first observe that advantage deviations from the average are higher in the multi-datacenter setup than same-datacenter setup, due to the heterogeneous consumer-provider latencies and random sampling of consumer locations from the dataset, whereas all consumers have 0ms network latency to all providers in the same-datacenter setup.
Under a cuckoo-delay attack, ``DeSearch-like'' and ``COoL w/o TT'' in \Cref{fig:byz-multiDC-cs-att} behave similarly to ``COoL w/o TT'' in the same-datacenter setup (\Cref{fig:byz-sameDC-cs-att-TT}), but with an accentuated malicious advantage even for low malicious presence.
In comparison, \Cref{fig:byz-multiDC-cs-att}'s ``COoL w/ TT'' remains under 7\% additional malicious share for $p_{M}\leq p_{exodus}^{delay}$.
This behavior is similar to its equivalent in the same-datacenter setup.
In other words, even in the multi-datacenter setup, \sysname \emph{with} Trusted Time under the powerful two-sided attack exhibits a resilience close to that of \sysname \emph{without} Trusted Time under the weaker provider-side-only delay attack.

\section{Related work}\label{sec:related_work}

\sysname~addresses issues at the intersection of multiple research domains, which we compare here.

\subsubsection{Decentralized search mechanisms}
Few recent works focus on enabling a more decentralized governance over search mechanisms. 
DHT-based solutions like HypeerCube~\cite{zichichi_towards_2021} and Ditto~\cite{keizerDittoDecentralisedSimilarity2023} do not guarantee result completeness. 
TheGraph~\cite{Graph} indexes Ethereum smart contracts but lacks completeness guarantees. 
DeSearch~\cite{liBringingDecentralizedSearch2021} uses distributed TEEs for subsecond latencies and result integrity, but is vulnerable to IHS attacks due to the host's control over the network interfaces.
While Kadabra~\cite{zhangKadabraAdaptingKademlia2023a} does not rely on TEEs and instead proposes a DHT-based search system based on Kademlia~\cite{maymounkovKademliaPeerPeerInformation2002}, it does not address IHS and suffers from multi-second latencies.

\subsubsection{Reputation systems}
\sysname's provider selection policy can be seen as a reputation system, 
locally updated by each consumer while only cooperating with TEEs, 
i.e., trustlessly with respect to other protocol actors. 
More advanced reputation systems open many attack vectors, notably well-studied in the peer-to-peer system literature~\cite{almasoudSmartContractsBlockchainbased2020,belliniBlockchainBasedDistributedTrust2020,hasanPrivacyPreservingReputationSystems2022,pereiraReputationSystemsFramework2023}.

\subsubsection{Decentralized QoS-based provider selection}
Provider selection protocols address IHS indirectly but usually assume a non-adversarial setting. 
\sysname\space adapts these mechanisms to malicious environments using TEEs. 
Go-with-the-winner~\cite{liuGowiththewinnerPerformanceBased2016} selects providers based on past round-trip latencies in a non-adversarial setting, 
while sPOT~\cite{panigrahyAnalysisEvaluationProximitybased2022} uses proximity-based selection, which we have shown to be vulnerable to IHS. 
DONAR~\cite{wendellDONARDecentralizedServer2010} introduces a mapper role to select providers, but it can be bypassed by malicious consumers, raising IHS risk.

\subsubsection{Resilient time measurements with TEEs}

This work focuses on leveraging Intel SGX TEEs, readily available at multiple major Cloud Service Providers. 
The \texttt{sgx\_get\_trusted\_time} primitive is unavailable for Linux since the 2020 SGX SDK's version 2.9, leading previous trusted time solutions to be deprecated~\cite{topleVeriCountVerifiableResource2018,anwarSecuringTimeUntrusted2019} and new ones to be proposed~\cite{hamidyT3EPracticalSolution2023,fernandezTriadTrustedTimestamps2023a}. 
T3E~\cite{hamidyT3EPracticalSolution2023} uses a local TPM as a source of trusted time for the TEE who continually polls it for timestamps.
However, the latency for a usable timestamp in the client execution is TPM-specific, in the order of tens to hundreds of milliseconds.
Moreover, T3E relies on remote users to detect a throughput drop when the host performs a delay attack between the TPM and the TEE, which is complex to implement in practice in our context, with many concurrent clients instead of a single one commissioning a task.
Triad~\cite{fernandezTriadTrustedTimestamps2023a} uses clusters of multiple TEEs in the same datacenter to maintain a shared notion of elapsed time. 
New SGX processors additionally propose the AEX-Notify functionality~\cite{constableAEXNotifyThwartingPrecise2023}, which enables developpers to implement arbitrary code to run when a TEE enclave resumes its execution after an interruption.
This interrupt-awareness enables Triad to then refresh the local timestamp using either in-cluster TEEs or a remote time authority.
Because time is measured inside the TEE with Triad, it offers a mean clock error in the hundreds of microseconds, within our requirements for \sysname.
% T-Lease~\cite{trachTLeaseTrustedLease2020a} is similar to Triad's local timekeeping protocol, using standalone TEEs to measure lower bounds on elapsed delays, before AEX-Notify was proposed.
% S-FaaS~\cite{alder2019s} measures delays in CPU cycles or instructions instead wall clock time, using Intel Transactional Synchronization eXtensions (TSX). 
Alder et al.~\cite{alderTimeChallengesTemporal2023} formalize security levels for TEE timers and survey recent solutions for the main TEE architectures, including Intel SGX and TDX, AMD SEV, as well as ARM TrustZone and CCA.
\section{Conclusion}\label{sec:conclusion}

In this paper, we highlight how providers of a marketplace's distributed search service are able to give favorable treatment to part of the userbase, through \emph{information head-start} attacks, even if Trusted Execution Environments (TEEs) are used to guarantee the integrity and confidentiality of the search mechanism.

We show that quality-of-service-aware provider selection as performed by our proposed mechanism \sysname, on top of TEEs, is paramount to counter such malicious provider behavior. 
Thanks to \sysname, we show that whether users collude or not with malicious providers, they do not get an edge over others in terms of fresher market state information, as long as there is enough honest provider throughput to serve all honest requests.
%Future research directions include investigating the dynamic spawning or retirement of service providers over the network topology, tailored to the users' latency requirements to deal with the intrinsic heterogeneity of client latency distributions.

\bibliographystyle{ieeetran}
\bibliography{imports/bibliography}

\begin{thebibliography}{10}
\providecommand{\url}[1]{#1}
\csname url@samestyle\endcsname
\providecommand{\newblock}{\relax}
\providecommand{\bibinfo}[2]{#2}
\providecommand{\BIBentrySTDinterwordspacing}{\spaceskip=0pt\relax}
\providecommand{\BIBentryALTinterwordstretchfactor}{4}
\providecommand{\BIBentryALTinterwordspacing}{\spaceskip=\fontdimen2\font plus
\BIBentryALTinterwordstretchfactor\fontdimen3\font minus
  \fontdimen4\font\relax}
\providecommand{\BIBforeignlanguage}[2]{{%
\expandafter\ifx\csname l@#1\endcsname\relax
\typeout{** WARNING: IEEEtran.bst: No hyphenation pattern has been}%
\typeout{** loaded for the language `#1'. Using the pattern for}%
\typeout{** the default language instead.}%
\else
\language=\csname l@#1\endcsname
\fi
#2}}
\providecommand{\BIBdecl}{\relax}
\BIBdecl

\bibitem{skim}
\BIBentryALTinterwordspacing
J.~Greig, ``Europol identifies hundreds of e-commerce platforms used in digital
  skimming attacks,'' 2023. [Online]. Available:
  \url{https://therecord.media/europol-identifies-hundreds-ecommerce-skimmers}
\BIBentrySTDinterwordspacing

\bibitem{cuthbertsonFacebookUsersReport2021}
\BIBentryALTinterwordspacing
A.~Cuthbertson, ``Facebook down: {{Users}} report issues with {{Messenger}} and
  {{Instagram}},'' 2021. [Online]. Available:
  \url{https://www.independent.co.uk/tech/facebook-down-messenger-instagram-not-working-b1950938.html}
\BIBentrySTDinterwordspacing

\bibitem{swearingenWhenAmazonWeb2018}
\BIBentryALTinterwordspacing
J.~Swearingen, ``When {{Amazon Web Services Goes Down}}, {{So Does}} a {{Lot}}
  of the {{Web}},'' 2018. [Online]. Available:
  \url{https://nymag.com/intelligencer/2018/03/when-amazon-web-services-goes-down-so-does-a-lot-of-the-web.html}
\BIBentrySTDinterwordspacing

\bibitem{CensorshipGoogle2024}
\BIBentryALTinterwordspacing
``Censorship by {{Google}},'' 2024. [Online]. Available:
  \url{https://en.wikipedia.org/w/index.php?title=Censorship_by_Google&oldid=1220502028}
\BIBentrySTDinterwordspacing

\bibitem{gargSteemitCensoringUsers2019}
\BIBentryALTinterwordspacing
P.~Garg, ``Steemit {{Censoring Users}} on '{{Immutable}}' {{Social Media
  Blockchain}},'' 2019. [Online]. Available:
  \url{https://cryptoslate.com/steemit-censoring-users-immutable-blockchain-social-media/}
\BIBentrySTDinterwordspacing

\bibitem{glaserHowAppleAmazon2017}
\BIBentryALTinterwordspacing
A.~Glaser, ``How {{Apple}} and {{Amazon Are Aiding Chinese Censors}},'' 2017.
  [Online]. Available:
  \url{https://slate.com/technology/2017/08/apple-and-amazon-are-helping-china-censor-the-internet.html}
\BIBentrySTDinterwordspacing

\bibitem{liBringingDecentralizedSearch2021}
M.~Li, J.~Zhu, T.~Zhang, C.~Tan, Y.~Xia, S.~Angel, and H.~Chen,
  ``\BIBforeignlanguage{en}{Bringing {Decentralized} {Search} to
  {Decentralized} {Services}},'' in \emph{\BIBforeignlanguage{en}{15th {USENIX}
  Symposium on Operating Systems Design and Implementation ({OSDI} 21)}}, 2021,
  pp. 331--347.

\bibitem{keizerDittoDecentralisedSimilarity2023}
N.~V. Keizer, O.~Ascigil, M.~Kr{\'o}l, and G.~Pavlou, ``Ditto: Towards
  decentralised similarity search for web3 services,'' in \emph{2023 IEEE
  International Conference on Decentralized Applications and Infrastructures
  (DAPPS)}.\hskip 1em plus 0.5em minus 0.4em\relax IEEE, 2023, pp. 66--75.

\bibitem{zichichi_towards_2021}
M.~Zichichi, L.~Serena, S.~Ferretti, and G.~D’Angelo, ``Towards
  {Decentralized} {Complex} {Queries} over {Distributed} {Ledgers}: a {Data}
  {Marketplace} {Use}-case,'' in \emph{2021 {International} {Conference} on
  {Computer} {Communications} and {Networks} ({ICCCN})}, Jul. 2021, pp. 1--6.

\bibitem{eskandariSoKTransparentDishonesty2020}
S.~Eskandari, S.~Moosavi, and J.~Clark, ``{{SoK}}: {{Transparent Dishonesty}}:
  {{Front-Running Attacks}} on {{Blockchain}},'' in \emph{Financial
  {{Cryptography}} and {{Data Security}}}, A.~Bracciali, J.~Clark, F.~Pintore,
  P.~B. Rønne, and M.~Sala, Eds.\hskip 1em plus 0.5em minus 0.4em\relax
  Springer International Publishing, 2020, pp. 170--189.

\bibitem{baumSoKMitigationFrontRunning2022}
\BIBentryALTinterwordspacing
C.~Baum, J.~H.-y. Chiang, B.~David, T.~K. Frederiksen, and L.~Gentile,
  ``{{SoK}}: {{Mitigation}} of {{Front-Running}} in {{Decentralized
  Finance}},'' in \emph{Financial {{Cryptography}} and {{Data Security}}.
  {{FC}} 2022 {{International Workshops}} - {{CoDecFin}}, {{DeFi}}, {{Voting}},
  {{WTSC}}, {{Grenada}}, {{May}} 6, 2022, {{Revised Selected Papers}}}, ser.
  Lecture {{Notes}} in {{Computer Science}}, S.~Matsuo, L.~Gudgeon,
  A.~Klages-Mundt, D.~P. Hernandez, S.~Werner, T.~Haines, A.~Essex,
  A.~Bracciali, and M.~Sala, Eds., vol. 13412.\hskip 1em plus 0.5em minus
  0.4em\relax Springer, 2022, pp. 250--271. [Online]. Available:
  \url{https://doi.org/10.1007/978-3-031-32415-4\_17}
\BIBentrySTDinterwordspacing

\bibitem{hamidyT3EPracticalSolution2023}
G.~M. Hamidy, P.~Philippaerts, and W.~Joosen, ``{{T3E}}: {{A Practical
  Solution}} to~{{Trusted Time}} in~{{Secure Enclaves}},'' in \emph{Network and
  {{System Security}}}, ser. Lecture {{Notes}} in {{Computer Science}}, S.~Li,
  M.~Manulis, and A.~Miyaji, Eds.\hskip 1em plus 0.5em minus 0.4em\relax
  Springer Nature Switzerland, 2023, pp. 305--326.

\bibitem{fernandezTriadTrustedTimestamps2023a}
\BIBentryALTinterwordspacing
G.~Fernandez, A.~Brito, and C.~Fetzer, ``Triad: {{Trusted Timestamps}} in
  {{Untrusted Environments}},'' in \emph{2023 {{IEEE International Conference}}
  on {{Cloud Computing Technology}} and {{Science}} ({{CloudCom}})}, 2023, pp.
  169--176. [Online]. Available:
  \url{https://ieeexplore.ieee.org/abstract/document/10475818}
\BIBentrySTDinterwordspacing

\bibitem{coolTEEcode}
M.~Bettinger, ``{COoL-TEE} source code,''
  \url{https://github.com/RedChainLab/COoL-TEE}, 2024.

\bibitem{panigrahyAnalysisEvaluationProximitybased2022}
N.~K. Panigrahy, T.~Vasantam, P.~Basu, D.~Towsley, A.~Swami, and K.~K. Leung,
  ``On the analysis and evaluation of proximity-based load-balancing
  policies,'' \emph{ACM Transactions on Modeling and Performance Evaluation of
  Computing Systems}, vol.~7, no. 2-4, pp. 1--27, 2022.

\bibitem{stathakopoulouAddingFairnessOrder2021}
C.~Stathakopoulou, S.~R{\"u}sch, M.~Brandenburger, and M.~Vukoli{\'c}, ``Adding
  fairness to order: Preventing front-running attacks in bft protocols using
  tees,'' in \emph{2021 40th International Symposium on Reliable Distributed
  Systems (SRDS)}.\hskip 1em plus 0.5em minus 0.4em\relax IEEE, 2021, pp.
  34--45.

\bibitem{kelkarThemisFastStrong2023}
\BIBentryALTinterwordspacing
M.~Kelkar, S.~Deb, S.~Long, A.~Juels, and S.~Kannan, ``Themis: {{Fast}},
  {{Strong Order-Fairness}} in {{Byzantine Consensus}},'' in \emph{Proceedings
  of the 2023 {{ACM SIGSAC Conference}} on {{Computer}} and {{Communications
  Security}}}, ser. {{CCS}} '23.\hskip 1em plus 0.5em minus 0.4em\relax
  Association for Computing Machinery, 2023, pp. 475--489. [Online]. Available:
  \url{https://dl.acm.org/doi/10.1145/3576915.3616658}
\BIBentrySTDinterwordspacing

\bibitem{costan2016intel}
\BIBentryALTinterwordspacing
V.~Costan and S.~Devadas, ``Intel {{SGX}} explained,'' \emph{Cryptology ePrint
  Archive}, 2016. [Online]. Available: \url{https://eprint.iacr.org/2016/086}
\BIBentrySTDinterwordspacing

\bibitem{wangCircuitORAMTightness2015}
\BIBentryALTinterwordspacing
X.~Wang, H.~Chan, and E.~Shi, ``Circuit {{ORAM}}: {{On Tightness}} of the
  {{Goldreich-Ostrovsky Lower Bound}},'' in \emph{Proceedings of the 22nd {{ACM
  SIGSAC Conference}} on {{Computer}} and {{Communications Security}}}, ser.
  {{CCS}} '15.\hskip 1em plus 0.5em minus 0.4em\relax Association for Computing
  Machinery, 2015, pp. 850--861. [Online]. Available:
  \url{https://doi.org/10.1145/2810103.2813634}
\BIBentrySTDinterwordspacing

\bibitem{TrustedTimeMonotonic}
\BIBentryALTinterwordspacing
S.~Cen and B.~Zhang. (2019) Trusted {{Time}} and {{Monotonic Counters}} with
  {{Intel}}® {{Software Guard Extensions Platform Services}}. Intel. [Online].
  Available:
  \url{https://www.intel.com/content/www/us/en/content-details/671564/trusted-time-and-monotonic-counters-with-intel-software-guard-extensions-platform-services.html}
\BIBentrySTDinterwordspacing

\bibitem{alderTimeChallengesTemporal2023}
\BIBentryALTinterwordspacing
F.~Alder, G.~Scopelliti, J.~Van~Bulck, and J.~T. Mühlberg, ``About {{Time}}:
  {{On}} the {{Challenges}} of {{Temporal Guarantees}} in {{Untrusted
  Environments}},'' in \emph{Proceedings of the 6th {{Workshop}} on {{System
  Software}} for {{Trusted Execution}}}, ser. {{SysTEX}} '23.\hskip 1em plus
  0.5em minus 0.4em\relax Association for Computing Machinery, 2023, pp.
  27--33. [Online]. Available: \url{https://doi.org/10.1145/3578359.3593038}
\BIBentrySTDinterwordspacing

\bibitem{constableAEXNotifyThwartingPrecise2023}
\BIBentryALTinterwordspacing
S.~Constable, J.~V. Bulck, X.~Cheng, Y.~Xiao, C.~Xing, I.~Alexandrovich,
  T.~Kim, F.~Piessens, M.~Vij, and M.~Silberstein,
  ``\{\vphantom\}{{AEX-Notify}}\vphantom\{\}: {{Thwarting Precise}}
  \{\vphantom\}{{Single-Stepping}}\vphantom\{\} {{Attacks}} through {{Interrupt
  Awareness}} for {{Intel}} \{\vphantom\}{{SGX}}\vphantom\{\} {{Enclaves}},''
  2023, pp. 4051--4068. [Online]. Available:
  \url{https://www.usenix.org/conference/usenixsecurity23/presentation/constable}
\BIBentrySTDinterwordspacing

\bibitem{raymondNtpsecSecureHardened2016a}
E.~S. Raymond, ``Ntpsec: A secure, hardened {{NTP}} implementation,''
  \emph{Linux Journal}, vol. 2016, no. 270, 2016.

\bibitem{ClockDisciplineAlgorithm}
\BIBentryALTinterwordspacing
{Network Time Foundation}. (2022) Clock filter and discipline algorithms. NTP:
  Network Time Protocol. [Online]. Available:
  \url{https://www.ntp.org/documentation/4.2.8-series/{filter,discipline}/}
\BIBentrySTDinterwordspacing

\bibitem{slivkins2019introduction}
A.~Slivkins \emph{et~al.}, ``Introduction to multi-armed bandits,''
  \emph{Foundations and Trends{\textregistered} in Machine Learning}, vol.~12,
  no. 1-2, pp. 1--286, 2019.

\bibitem{fiducioso2019safe}
M.~Fiducioso, S.~Curi, B.~Schumacher, M.~Gwerder, and A.~Krause, ``Safe
  contextual bayesian optimization for sustainable room temperature pid control
  tuning,'' \emph{arXiv preprint arXiv:1906.12086}, 2019.

\bibitem{xu2023config}
W.~Xu, Y.~Jiang, B.~Svetozarevic, P.~Heer, and C.~N. Jones, ``Config:
  Constrained efficient global optimization for closed-loop control system
  optimization with unmodeled constraints,'' \emph{IFAC-PapersOnLine}, vol.~56,
  no.~2, pp. 513--518, 2023.

\bibitem{auer2002nonstochastic}
P.~Auer, N.~Cesa-Bianchi, Y.~Freund, and R.~E. Schapire, ``The nonstochastic
  multiarmed bandit problem,'' \emph{SIAM journal on computing}, vol.~32,
  no.~1, pp. 48--77, 2002.

\bibitem{vargaOverviewOMNeTSimulation2008}
A.~Varga and R.~Hornig, ``An overview of the omnet++ simulation environment,''
  in \emph{1st International ICST Conference on Simulation Tools and Techniques
  for Communications, Networks and Systems}, 2010.

\bibitem{charyyevLatencyComparisonCloud2020}
\BIBentryALTinterwordspacing
B.~Charyyev, E.~Arslan, and M.~H. Gunes, ``Latency {{Comparison}} of {{Cloud
  Datacenters}} and {{Edge Servers}},'' in \emph{{{GLOBECOM}} 2020 - 2020
  {{IEEE Global Communications Conference}}}, 2020, pp. 1--6. [Online].
  Available: \url{https://ieeexplore.ieee.org/document/9322406}
\BIBentrySTDinterwordspacing

\bibitem{RIPEAtlasRIPE}
\BIBentryALTinterwordspacing
{{RIPE Atlas}} - {{RIPE Network Coordination Centre}}. [Online]. Available:
  \url{https://atlas.ripe.net/}
\BIBentrySTDinterwordspacing

\bibitem{mitzenmacherPowerTwoChoices2001}
M.~Mitzenmacher, ``The power of two choices in randomized load balancing,''
  \emph{IEEE Transactions on Parallel and Distributed Systems}, vol.~12,
  no.~10, pp. 1094--1104, 2001.

\bibitem{HorizontalPodAutoscaling}
\BIBentryALTinterwordspacing
{Kubernetes}. (2024) Horizontal {{Pod Autoscaling}}. [Online]. Available:
  \url{https://kubernetes.io/docs/tasks/run-application/horizontal-pod-autoscale/}
\BIBentrySTDinterwordspacing

\bibitem{Graph}
{The Graph}, 2018, website available at: https://thegraph.com/.

\bibitem{zhangKadabraAdaptingKademlia2023a}
\BIBentryALTinterwordspacing
Y.~Zhang and S.~B. Venkatakrishnan, ``Kadabra: {{Adapting Kademlia}} for the
  {{Decentralized Web}},'' in \emph{Financial {{Cryptography}} and {{Data
  Security}} - 27th {{International Conference}}, {{FC}} 2023, {{Bol}},
  {{Brač}}, {{Croatia}}, {{May}} 1-5, 2023, {{Revised Selected Papers}},
  {{Part II}}}, ser. Lecture {{Notes}} in {{Computer Science}}, F.~Baldimtsi
  and C.~Cachin, Eds., vol. 13951.\hskip 1em plus 0.5em minus 0.4em\relax
  Springer, 2023, pp. 327--345. [Online]. Available:
  \url{https://doi.org/10.1007/978-3-031-47751-5\_19}
\BIBentrySTDinterwordspacing

\bibitem{maymounkovKademliaPeerPeerInformation2002}
P.~Maymounkov and D.~Mazières, ``Kademlia: {{A Peer-to-Peer Information System
  Based}} on the {{XOR Metric}},'' in \emph{Peer-to-{{Peer Systems}}},
  P.~Druschel, F.~Kaashoek, and A.~Rowstron, Eds.\hskip 1em plus 0.5em minus
  0.4em\relax Springer, 2002, pp. 53--65.

\bibitem{almasoudSmartContractsBlockchainbased2020}
\BIBentryALTinterwordspacing
A.~S. Almasoud, F.~K. Hussain, and O.~K. Hussain, ``Smart contracts for
  blockchain-based reputation systems: {{A}} systematic literature review,''
  vol. 170, p. 102814, 2020. [Online]. Available:
  \url{https://www.sciencedirect.com/science/article/pii/S108480452030285X}
\BIBentrySTDinterwordspacing

\bibitem{belliniBlockchainBasedDistributedTrust2020}
\BIBentryALTinterwordspacing
E.~Bellini, Y.~Iraqi, and E.~Damiani, ``Blockchain-{{Based Distributed Trust}}
  and {{Reputation Management Systems}}: {{A Survey}},'' vol.~8, pp.
  21\,127--21\,151, 2020. [Online]. Available:
  \url{https://ieeexplore.ieee.org/document/8970496/?arnumber=8970496}
\BIBentrySTDinterwordspacing

\bibitem{hasanPrivacyPreservingReputationSystems2022}
\BIBentryALTinterwordspacing
O.~Hasan, L.~Brunie, and E.~Bertino, ``Privacy-{{Preserving Reputation Systems
  Based}} on {{Blockchain}} and {{Other Cryptographic Building Blocks}}: {{A
  Survey}},'' vol.~55, no.~2, pp. 32:1--32:37, 2022. [Online]. Available:
  \url{https://doi.org/10.1145/3490236}
\BIBentrySTDinterwordspacing

\bibitem{pereiraReputationSystemsFramework2023}
\BIBentryALTinterwordspacing
R.~H. Pereira, M.~J. Gonçalves, and M.~A.~G. Magalhães, ``Reputation
  {{Systems}}: {{A}} framework for attacks and frauds classification,'' vol.~8,
  no.~1, p. 19218, 2023. [Online]. Available:
  \url{https://www.jisem-journal.com/article/reputation-systems-a-framework-for-attacks-and-frauds-classification-12830}
\BIBentrySTDinterwordspacing

\bibitem{liuGowiththewinnerPerformanceBased2016}
C.~Liu, R.~K. Sitaraman, and D.~Towsley, ``Go-with-the-winner: Performance
  based client-side server selection,'' in \emph{2016 IFIP Networking
  Conference (IFIP Networking) and Workshops}.\hskip 1em plus 0.5em minus
  0.4em\relax IEEE, 2016, pp. 404--412.

\bibitem{wendellDONARDecentralizedServer2010}
P.~Wendell, J.~W. Jiang, M.~J. Freedman, and J.~Rexford, ``Donar: decentralized
  server selection for cloud services,'' in \emph{Proceedings of the ACM
  SIGCOMM 2010 conference}, 2010, pp. 231--242.

\bibitem{topleVeriCountVerifiableResource2018}
S.~Tople, S.~Park, M.~S. Kang, and P.~Saxena, ``{{VeriCount}}: {{Verifiable
  Resource Accounting Using Hardware}} and {{Software Isolation}},'' in
  \emph{Applied {{Cryptography}} and {{Network Security}}}, B.~Preneel and
  F.~Vercauteren, Eds.\hskip 1em plus 0.5em minus 0.4em\relax Springer
  International Publishing, 2018, pp. 657--677.

\bibitem{anwarSecuringTimeUntrusted2019}
\BIBentryALTinterwordspacing
F.~M. Anwar, L.~Garcia, X.~Han, and M.~Srivastava, ``Securing {{Time}} in
  {{Untrusted Operating Systems}} with {{TimeSeal}},'' in \emph{2019 {{IEEE
  Real-Time Systems Symposium}} ({{RTSS}})}, 2019, pp. 80--92. [Online].
  Available: \url{https://ieeexplore.ieee.org/abstract/document/9052115}
\BIBentrySTDinterwordspacing

\end{thebibliography}

\end{document}